\documentclass[%
 reprint,
superscriptaddress,
amsmath,amssymb,
aps,
prb,
]{revtex4-1}
\usepackage[T1]{fontenc}
\usepackage{graphicx}
\usepackage{dcolumn}
\usepackage{bm}
\usepackage{amsmath}
\usepackage{lmodern}
\usepackage{mathtools}
\usepackage[colorlinks,citecolor=black,linkcolor=black,urlcolor=black]{hyperref}

\begin{document}

\title{Angular dependence of giant Zeeman effect for semi-magnetic cavity polaritons}
\author{R.~Mirek}
\email{rafal.mirek@student.uw.edu.pl}
\author{M.~Kr\'ol}
\author{K.~Lekenta}
\author{J.-G.~Rousset}
\author{M.~Nawrocki}
\affiliation{%
 Institute of Experimental Physics, Faculty of Physics,\\
 University of Warsaw, ul. Pasteura 5, PL-02-093 Warsaw, Poland}
\author{M.~Kulczykowski}
\author{M.~Matuszewski}
\affiliation{Institute of Physics, Polish Academy of Sciences, al.
Lotnik\'{o}w 32/46, PL-02-668 Warsaw, Poland}
\author{J.~Szczytko}
\author{W.~Pacuski}
\author{B.~Pi\k{e}tka}
\email{barbara.pietka@fuw.edu.pl}

\affiliation{%
 Institute of Experimental Physics, Faculty of Physics,\\
 University of Warsaw, ul. Pasteura 5, PL-02-093 Warsaw, Poland}

\begin{abstract}
The observation of spin-related phenomena of microcavity polaritons has been limited due to weak Zeeman effect of non-magnetic semiconductors. We demonstrate that the incorporation of magnetic ions into quantum wells placed in a non-magnetic microcavity results in enhanced effects of magnetic field on exciton-polaritons. We show that in such a structure the Zeeman splitting of exciton-polaritons strongly depends on the photon-exciton detuning and polariton wavevector. Our experimental data are explained by a model where the impact of magnetic field on the lower polariton state is
directly inherited from the excitonic component, and the coupling strength to cavity photon is modified by external magnetic field.
\end{abstract}
\maketitle

\section{\label{sec:intro}Introduction}

Cavity exciton-polaritons are quasiparticles resulting from the strong coupling between cavity optical modes and exciton resonances. The most frequently studied system consists of an exciton confined in a quantum well embedded in a semiconductor microcavity. During the last decade, intensive research showed that such a system is a well suited playground for
investigating fundamental objects like Bose-Einstein condensates \cite{Kasprzak, Balili} or polariton superfluids\cite{Amo_nature2009, nardin_natphys2011}, making the physics of non-equilibrium quantum fluids accessible at higher temperatures\cite{Levrat_PRB_2010, Klembt_PRL2015}. The unique features of polariton condensates set the basis for new polaritonic devices\cite{Liew_PhsE2011, Ballarini_Ncomms2013}. 

In recent years an interest in the magneto-optical properties of cavity polaritons has risen\cite{Kavokin_PRB_1997, Salis_PRB_2005,Solnyshkov_PRB2008, Solnyshkov_PRB2009, Larionov_PRL2010, Walker_PRL2011, Fisher_PRL2014, Pietka_PRB2015, Sturm_PRB2015, Kochereshko_SR_2016}. In the extensively studied GaAs-based
microcavities the polariton Zeeman splitting is of the order of the emission linewidth in magnetic fields easily accessible in laboratories \cite{Pietka_PRB2015}. However, semimagnetic semiconductors offer the opportunity to enhance magneto-optical effects via the exchange interaction between the $d$-shell electrons of a magnetic ion and the $s$-shell electrons and $p$-shell holes of the conduction and valence bands of the host material\cite{Gaj_PSSB1978}. This $s$,$p$-$d$ exchange interaction leads to enhanced magneto-optical effects like giant Faraday
rotation\cite{Gaj_SSCom1978} or giant Zeeman splitting\cite{Gaj_SScom1979}. At the same time, the incorporation of magnetic ions can lead also to photoluminescence (PL) quenching and an important lowering of optical quality, what is particularly striking in the case of Ga$_{1-x}$Mn$_x$As structures\cite{Szczytko_PRB_1999,Poggio2005}. Another type of material is a semiconductor system relatively resistant to manganese induced optical degradation: Cd$_{1-x}$Mn$_x$Te, which was already used as a high refractive index layer in microcavities with Cd$_{1-x}$Mn$_x$Te
quantum wells\cite{Ulmer_SLM97, Haddad_SSC_1999}. Indeed such a system exhibits strong magneto-optical properties\cite{Cubian_PRB_2003,Koba_JEWA2013,Koba_EPL2014} related to the shift of both exciton and photon states\cite{Ulmer_SLM97,Sadowski_PRB_1997},  therefore it can be used for tuning the exciton-photon energy difference and for tuning the Rabi oscillation frequency\cite{Brunetti_PSSC_2005, Brunetti_PRB2006, Brunetti_PRB2006_spin_dynamics}.

Our approach to semimagnetic cavity polaritons is based on redesigned structures where magnetic ions are inserted only in the quantum wells, while the cavity and the distributed Bragg reflectors (DBRs) are made of non-magnetic materials\cite{Pacuski_CGD2014,Rousset_JCG2014}. We show that in such a case the giant Zeeman effect of polaritons results only from the strong coupling of cavity photons with semimagnetic excitons confined in CdTe quantum wells containing manganese ions.

The giant Zeeman splitting of polaritons is studied in angle resolved photoluminescence and reflectivity experiments. A theoretical model taking into account the dependence of the splitting on the in plane wavevector and the photon-exciton detuning is described. These results are the first step towards the study of spinor polariton condensates\cite{Solnyshkov_PRB2009, Solnyshkov_PRB2012, Ohadi_PRX2015} with enhanced magneto-optical properties.

\section{\label{sec:sample} Sample and experimental setup}

\begin{figure}

 \includegraphics[width=6cm]{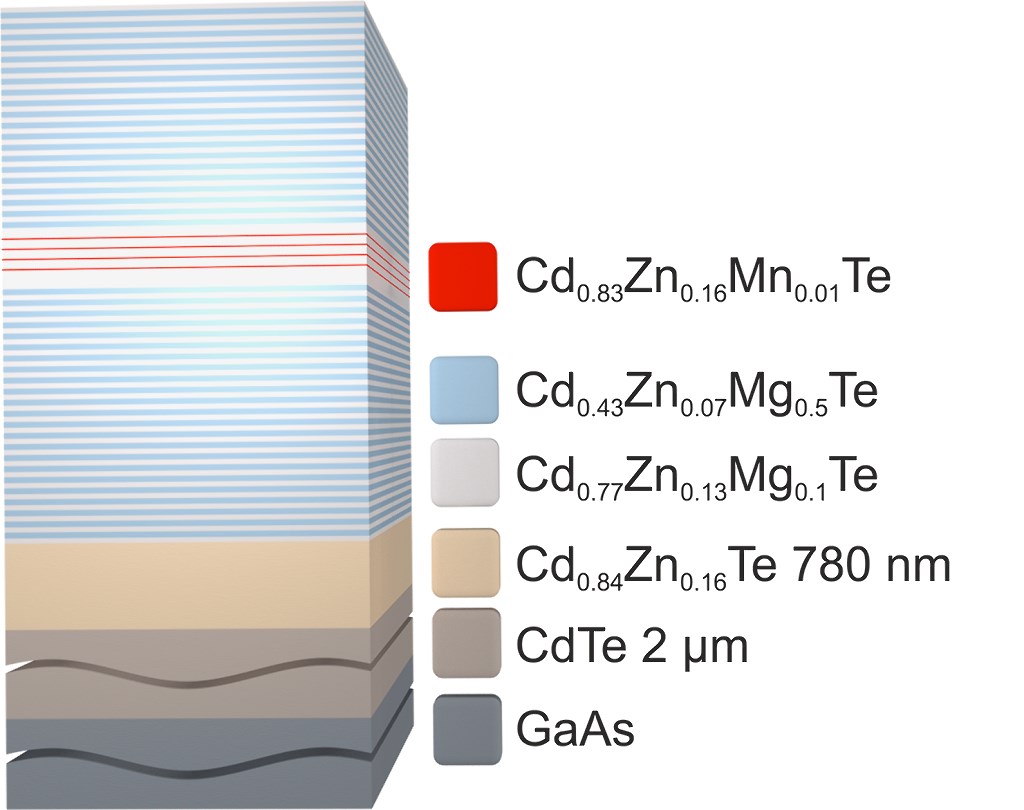}
 \caption{The structure of the microcavity sample: four 20 nm wide QWs (red) containing Mn$^{2+}$ ions sandwiched between 23 layers (bottom) and 20 layers (top) nonmagnetic distributed Bragg reflectors (alternating blue and white layers). Semimagnetic cavity polaritons result from the strong coupling of cavity photons with excitons confined in the semimagnetic QWs.}
 \label{fig:fig_1}
 \end{figure}

The sample studied in this work was grown using molecular beam epitaxy. On a (100) oriented GaAs substrate a 2~$\mu$m thick CdTe buffer is followed by a 780~nm thick Cd$_{0.86}$Zn$_{0.14}$Te buffer. Starting from this buffer layer, the whole structure is lattice matched to MgTe\cite{ Rou_JCG2013,Rousset_APL2015}. Photons are confined by distributed Bragg reflectors (DBRs) made of Cd$_{0.77}$Zn$_{0.13}$Mg$_{0.10}$Te for the high refractive index layers and Cd$_{0.43}$Zn$_{0.07}$Mg$_{0.50}$Te for the low refractive index layers. Respective refractive index\cite{Rou_JCG2013,Rousset_JCG2014} at the center wavelength $\lambda_0=735$~nm is $n_{high}=2.97$ and $n_{low}=2.61$. Four 20 nm wide Cd$_{0.83}$Zn$_{0.16}$Mn$_{0.01}$Te quantum wells (QWs) are placed at the antinodes of the electric field of the $3\lambda$ cavity. Since our structure is designed to enhance magneto-optical effects in cavity-polaritons only via the excitonic component, manganese is incorporated only in quantum wells, not in the surrounding material, nor in the DBRs. The detailed  scheme of the sample structure is shown in \hyperref[fig:fig_1]{FIG.~\ref{fig:fig_1}}.

The sample is placed in a magnetic field up to $5$~T in a Faraday configuration on the cold finger of an optical cryostat at the temperature of $10$~K. The sample is excited non-resonantly by a continuous-wave Ti:Sapphire laser which energy is tuned to the first reflectivity minimum of the cavity stop-band on the high energy side: $E_{exc}=1.76$~eV ($\lambda _{exc}=705$~nm). The angle resolved photoluminescence and reflectivity spectra not resolved in polarization are collected, giving a direct information about the polariton dispersion (energy-wavevector dependence), with the wavevector
scale being proportional to the angle at which light is emitted from the structure~\cite{Kasprzak}.

\section{\label{sec:experiment}Experimental results: imaging polariton
dispersion by photoluminescence and reflectivity}

\begin{figure*}
  \includegraphics[width=14cm]{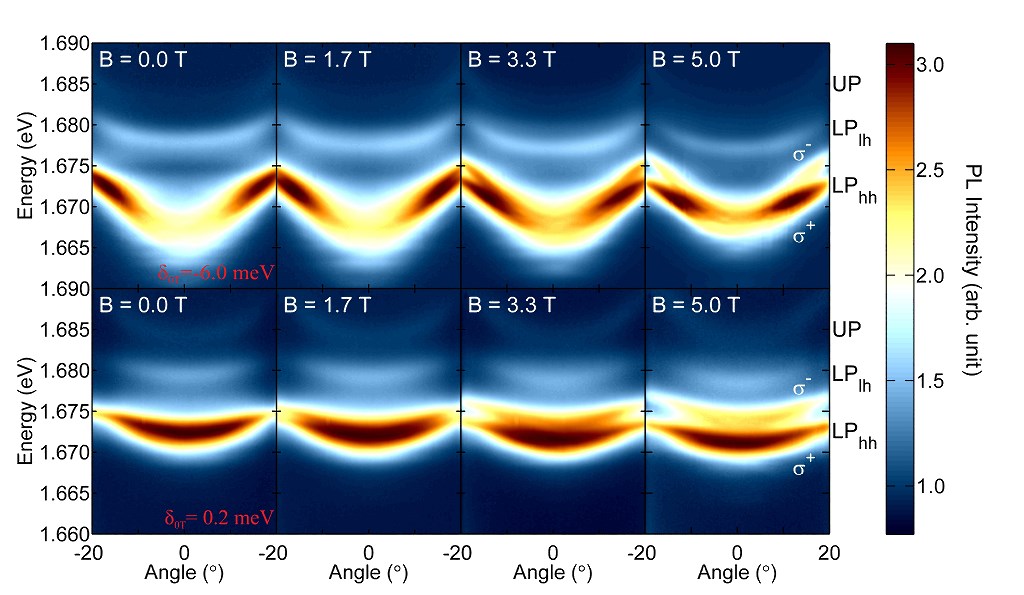}
  \caption{Angle resolved photoluminescence maps of semimagnetic exciton-polaritons for negative, $\delta_{0\textrm{T}}=-6.0$~meV, (top panel) and slightly positive, $\delta_{0\textrm{T}}=0.2$~meV, (bottom panel) photon-exciton detuning (determined at zero emission angle) at magnetic fields of: 0~T, 1.7~T, 3.3~T and 5.0~T. The giant Zeeman splitting is well resolved for positive detuning, but it is also observed for higher angles at negative detuning.}
  \label{fig:fig_2b_skala}
  \end{figure*}

\hyperref[fig:fig_2b_skala]{FIG.~\ref{fig:fig_2b_skala}} illustrates the evolution of the exciton-polariton dispersion in magnetic field for two representative cases of negative (top panel) and positive (bottom panel) photon-exciton detunings. The detuning is defined as the difference in energy between the bare cavity mode and the exciton energy at zero emission angle, $\delta$ = $E_{ph}-E_{exc}$, where $E_{ph}$, $E_{exc}$ are photon and exciton energies respectively. At magnetic field the detuning is defined separately for the two Zeeman-split polariton components. The maps show PL spectra measured as a function of the emission angle, for zero magnetic field (left side panels) and for magnetic fields increasing up to 5~T (right side panels). The Zeeman splitting of the lower polariton state increases with magnetic field, what is very well visible for positive detuning, where the excitonic contribution to the polariton wave function is dominant. Interestingly, also for negative detuning, i.e. photon like lower polariton states, a significant splitting can be observed at high magnetic field (upper-right panel). However in this
case, the splitting is visible for high emission angles. Since in a first approximation, the magnetic field affects only excitons, it does not change the photon states\cite{Kavokin_book2009,
Pietka_PRB2015}, we explain the magnetic-field evolution of a photon-like polariton, as a consequence of the exciton state contribution evolution with k-vector, magnetic field and detuning.

 \begin{figure*}
 \phantomsection
 \centering
 \includegraphics[width=13.5cm]{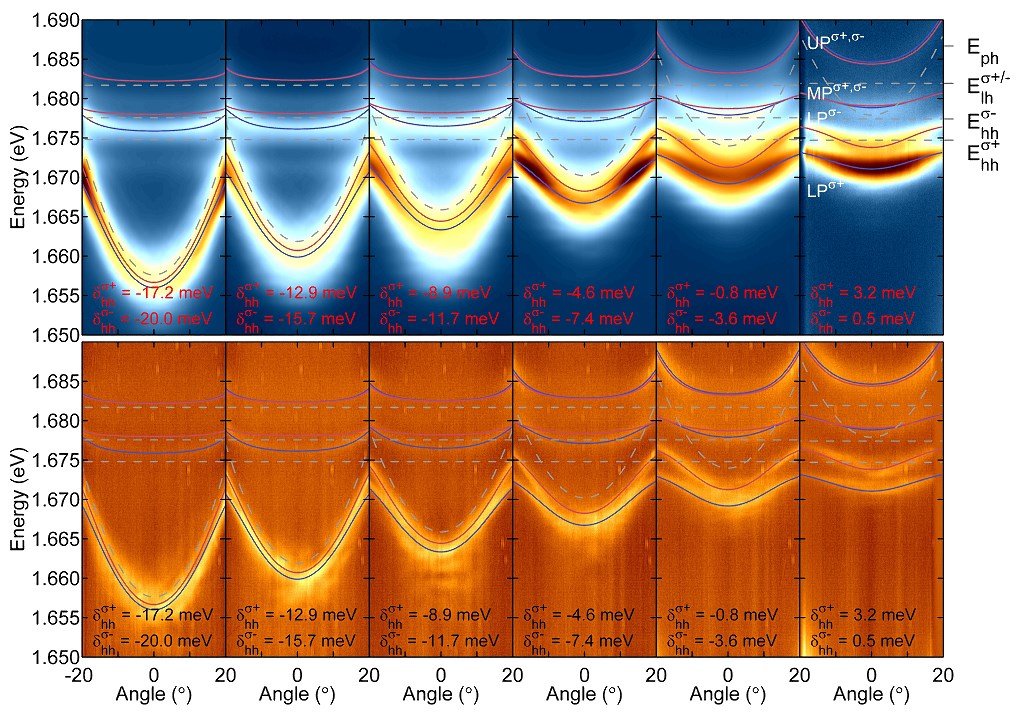}
 \caption{Set of angle resolved photoluminescence (top) and reflectivity (bottom) maps for different photon-exciton detuning (the detuning values are marked  in the images for $\sigma^+$ and $\sigma^-$ polarization) in a magnetic field of 5~T. Grey dashed curves figure the calculated energies of the uncoupled exciton and photon. The energies of polariton branches are marked by red ($\sigma^-$) and blue ($\sigma^+$) curves.}
 \label{fig:fig_3b_skala}
 \end{figure*}

A more detailed analysis of the impact of photon-exciton detuning on the polariton dispersion is presented in \hyperref[fig:fig_3b_skala]{FIG.~\ref{fig:fig_3b_skala}}, where sets of photoluminescence and reflectivity maps in a magnetic field of 5~T, measured for various values of photon-exciton detuning: from large negative (left side) to small positive (right side) detuning are presented. Both photoluminescence and reflectivity measurements lead to a consistent image of the polariton dispersion. The first conclusion to be imposed, is that the splitting is more visible for the lower polariton with a higher excitonic content. The splitting of photon-like polaritons (negative detuning) is visible only for large values of the wavevector and at zero wavevector, the Zeeman splitting is smaller than the linewidth. In the positive detuning case, the energy difference between the split components of the lower polariton is more pronounced. Furthermore it does not depend so strongly on the wavevector as in the case of negative detuning.

\section{\label{sec:model}Modelling polariton dispersion}

We derived a simple theoretical model to explain the behaviour of exciton-polaritons in magnetic field. Heavy holes excitons have a four-fold degeneracy of spin projection on the axis of the cavity: $\pm$2, $\pm$1. The excitons of a spin projection $\pm$2 remain dark, not active with respect to light coupling  or photon emission and are neglected in the following consideration. The optically active excitons, with +1 and $-$1 spin degeneracy, couple to $\sigma^+$ and $\sigma^-$ circularly polarized light, respectively, forming polaritons with two spin projections. Therefore in our model we consider two independent Hamiltonians for both circularly polarized subsystems, $\hat{H}^\pm$. Moreover, in our QWs the heavy and light hole separation energy is small, and we observe a strong contribution into the polariton states from excitons formed with light holes. In CdMnTe-based QWs the Zeeman splitting of the light hole excitons is one order of magnitude smaller than the splitting for heavy hole excitons\cite{Gaj_SScom1979}. Nevertheless we take into account the splitting of light holes in our model. Our coupling Hamiltonians take the following, three-level model form, each one corresponding to the coupling of heavy and light hole excitons to $\sigma^+$, $\sigma^-$ circular polarization of light:

\begin{equation}
\label{eq:Hamiltonian3D}
\hat{H}^\pm=\begin{pmatrix}
  E_{ph} & \frac{\hbar\Omega_{lh}^{\sigma\pm}}{2} &
\frac{\hbar\Omega_{hh}^{\sigma\pm}}{2}\\
  \frac{\hbar\Omega_{lh}^{\sigma\pm}}{2} & E_{lh}^{\sigma\pm} & 0\\
  \frac{\hbar\Omega_{hh}^{\sigma\pm}}{2} & 0 & E_{hh}^{\sigma\pm} \\
 \end{pmatrix}
\end{equation}
where $E_{ph}$ is the cavity photon energy and the energies of light hole and heavy hole excitons are $E_{lh}$ and $E_{hh}$, respectively. The coupling energy between the corresponding excitons with cavity photons is marked as $\Omega_{lh}$ and $\Omega_{hh}$. The symbols $\sigma^+$ and $\sigma^-$ correspond to two Zeeman-split components of excitons. The energy of both circularly polarized cavity photons, $E_{ph}$, is the same and it is expected to be independent of magnetic field, however, to verify this assumption we leave this parameter free in the fitting procedure of the model to the experimental data.

\begin{figure}
 \phantomsection
 \centering
 \includegraphics[width=8cm]{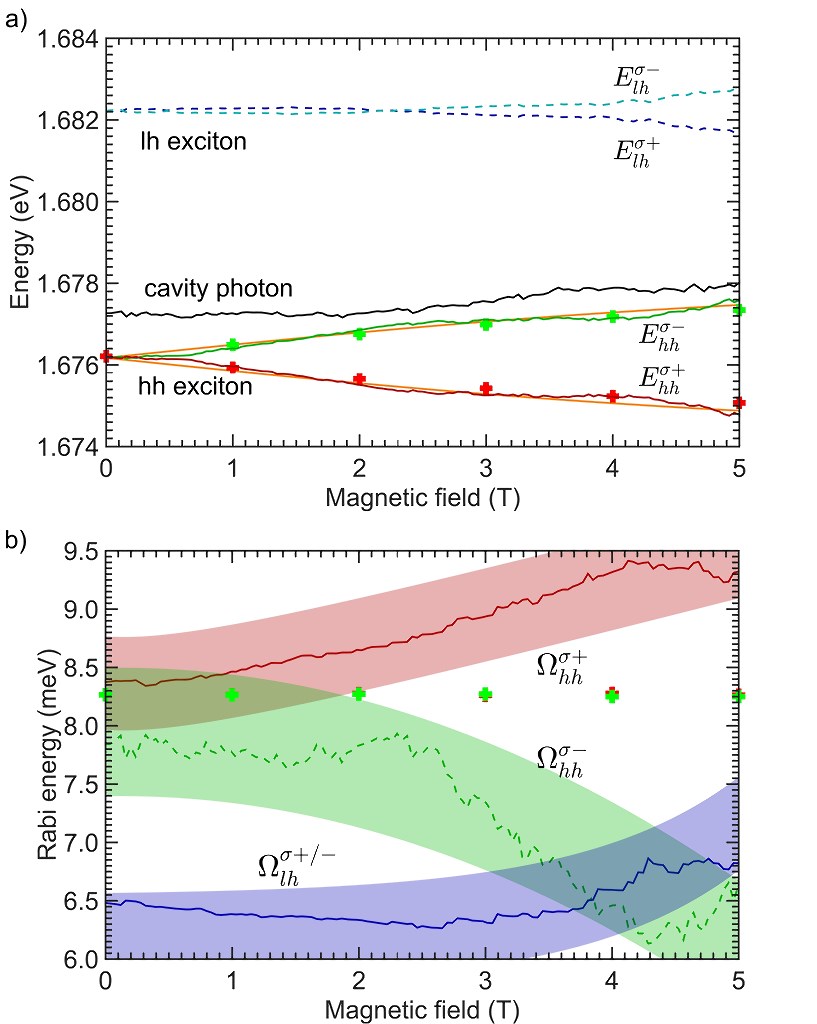}
 \caption{a) Energy change of heavy hole $E_{hh}$, light hole excitons $E_{lh}$ and a cavity photon induced by magnetic field. b)~Change of the Rabi energy in magnetic field. The experimental results are obtained from the fitting of a solution of the Schr\"{o}dinger equation with the hamiltonian (1) to the data illustrated in the last panel of FIG.~3. The shadow along the experimental curves represents the error margins of our fitting procedure. The dots illustrate the results of theoretical calculations based on a variational method applied to our structure.}
 \label{fig:fig_4c}
 \end{figure}

In the top panel of \hyperref[fig:fig_3b_skala]{FIG.~\ref{fig:fig_3b_skala}} we directly compare the experimental polariton dispersion curves with the theoretical model. Grey dashed curves illustrate the calculated energies of the bare photon and excitons. Red and blue curves give energies of the polaritons with $\sigma^-$ and $\sigma^+$ polarization. The fitting of the model to the photoluminescence experimental data allows to obtain the bare exciton and photon energies as well as the Rabi energy in magnetic field. \hyperref[fig:fig_4c]{FIG.~\ref{fig:fig_4c}} illustrates the fitted parameters for the data from the positive detuning (last PL map of \hyperref[fig:fig_3b_skala]{FIG.~\ref{fig:fig_3b_skala}}). \hyperref[fig:fig_4c]{FIG.~\ref{fig:fig_4c}} a) illustrates the energy change of bare photon and excitons in magnetic field. We observe that the photon energy does not change significantly in magnetic field, what confirms our previous assumption. In contrast, heavy hole exciton splits significantly in magnetic field.

In the mean field theory, the Zeeman splitting of a heavy hole exciton in semimagnetic semiconductors resulting from $s$,$p$-$d$ exchange interaction, scales with the magnetization of magnetic ions \cite{Gaj_PSSB1978}. Therefore, for low magnetic ion content, it is proportional to the Brillouin function $B_{5/2}$ describing magnetization of non-interacting Mn ions of spin 5/2 at the temperature of the experiment (10~K)~\cite{Gaj_SScom1979}. The observed almost linear dependence is due to the fact, that measurements were performed at the temperature of about 10~K what agrees with the shape of the Brillouin function at this temperature, marked with orange solid line in \hyperref[fig:fig_4c]{FIG.~\ref{fig:fig_4c}} a).

For better understanding of magnetic field evolution of electron and hole in our QW we calculated exciton energies by solving the full two-body Schr\"odinger problem using a variational method~\cite{Bastard_PRB1982,Ivchenko_PRB1992}. We assumed the exponential form of the relative electron-hole wavefunction and hyperbolic secant wavefunction for both electrons and holes in the direction perpendicular to the QW plane. We take into account change of the exciton energy by calculating the band offset of electron and hole as
\begin{equation*}
\Delta V^{\sigma_\pm}_{e}= \pm {N_{0}}\alpha x\left<S_z\right>,
\end{equation*}
\begin{equation}
\Delta V^{\sigma_\pm}_{hh}= \mp N_{0}\beta x\left<S_z\right>,
\end{equation}
where $\alpha$, $\beta$ are the exchange interaction constants between magnetic ions and the electrons and holes, respectively, and $1/N_{0}$ is the volume of the elementary crystal cell. We take into account bright excitons for which the spins of electrons and holes are aligned in opposite directions. The exchange integrals are taken according to~\cite{Gaj_SScom1979} as $N_{0}\alpha=220\,$meV and $N_{0}\beta=880\,$meV.
As the ion concentration is very low, the mean spin is taken in the dilute limit as
\begin{equation}
\left<S_z\right> = \frac{5}{2} B_{5/2} \left(5g\mu_B H/2k_{B}T \right).
\end{equation}
Here $\mu_B$ is the Bohr magneton, $g\approx 2$ is the g-factor of the ions, and $H$ is the strength of external magnetic field in the Faraday configuration. The exciton g-factor of nonmagnetic 20 nm wide QW\cite{Zhao1996} (g = 0.27) is small enough to be neglected. The theoretical energies shown in \hyperref[fig:fig_4c]{FIG.~\ref{fig:fig_4c}} a) with dots agree with experimental data when the sample background temperature $T=10$~K is used to estimate the mean spin polarization. Moreover, taking into account the increased overlap of the electron and hole wave functions in magnetic field and the energy shift of the electron and hole levels in magnetic field (therefore the different penetration of the wavefunction in the QW barrier), we calculated the effect of magnetic field on the exciton oscillator strength and on the exciton binding energy. We found that such magnetic field induced dependence is negligible in our structure, and consequently predicted Rabi energy (shown by dots in Fig. \hyperref[fig:fig_4c]{FIG.~\ref{fig:fig_4c}} b) is almost constant in magnetic field. It is in contrast to the previous study of (Cd,Mn)Te/(Cd,Mg)Te QWs~\cite{Ivchenko_PRB1992}, but the difference is fully explained by relatively low Mn concentration in the QW and the strength of the confining potential, which both makes that the perpendicular electron and hole wave-functions in our structure remain practically unaffected by the magnetic field. Prediction of almost constant Rabi energy is also different than in case of GaAs-based microcavities, where the Rabi energy and exciton oscillator strength increases in magnetic field\cite{Pietka_PRB2015}, but in II-VI semiconductors exciton size is much smaller and consequently direct effect of magnetic field is much weaker.

Experimentally, by fitting of the model described by Hamiltonians (1) to the observed emission lines, we observe that the Rabi energy of the heavy hole $\sigma^+$ polariton slightly increases in magnetic field and interestingly, the Rabi energy of the heavy hole $\sigma^-$ polariton is observed to decrease in magnetic field. The comparison of the results of the fitting with different parameters for constant and varied Rabi energy in magnetic field is illustrated in Supplemental Material (SI), where we demonstrate that the fitting to the experimental data is more accurate when the Rabi energy is a free parameter. The experimentally observed decrease of Rabi energy in $\Omega^{\sigma^-}_{hh}$ is attributed to the effect of field dependent inhomogenous broadening of the exciton linewidth, due to fluctuations of Mn distribution in the QWs. Such broadening is known for semimagnetic semiconductors~\cite{Sugakov_JPCM_2001,Komarov_JPCM_2006} and is manifested by a significant narrowing of the exciton emission linewidth in $\sigma^+$ polarization and broadening in $\sigma^-$ polarization.

We attribute the loss of Rabi energy to the gradual transition to weak coupling of the individual exciton modes. As the magnetically induced broadening is increased, a fraction of localized exciton modes moves out of resonance and becomes weakly coupled to photons, which leads to the decrease of the overall coupling strength. The observed Rabi energy is therefore in our case an effective parameter taking into account coupling of many excitonic transitions with slightly different energy to a cavity mode. The images illustrating the exciton-polariton branches including the inhomogenous broadening of the exciton line are presented in the SI. Such interpretation suggests that the value of the intrinsic Rabi coupling of individual exciton modes does not depend significantly on magnetic field, but due to the effect of inhomogenous Mn distribution in QW, the effective Rabi energy changes.

 \begin{figure}
 \centering
 \includegraphics[width=\textwidth/2]{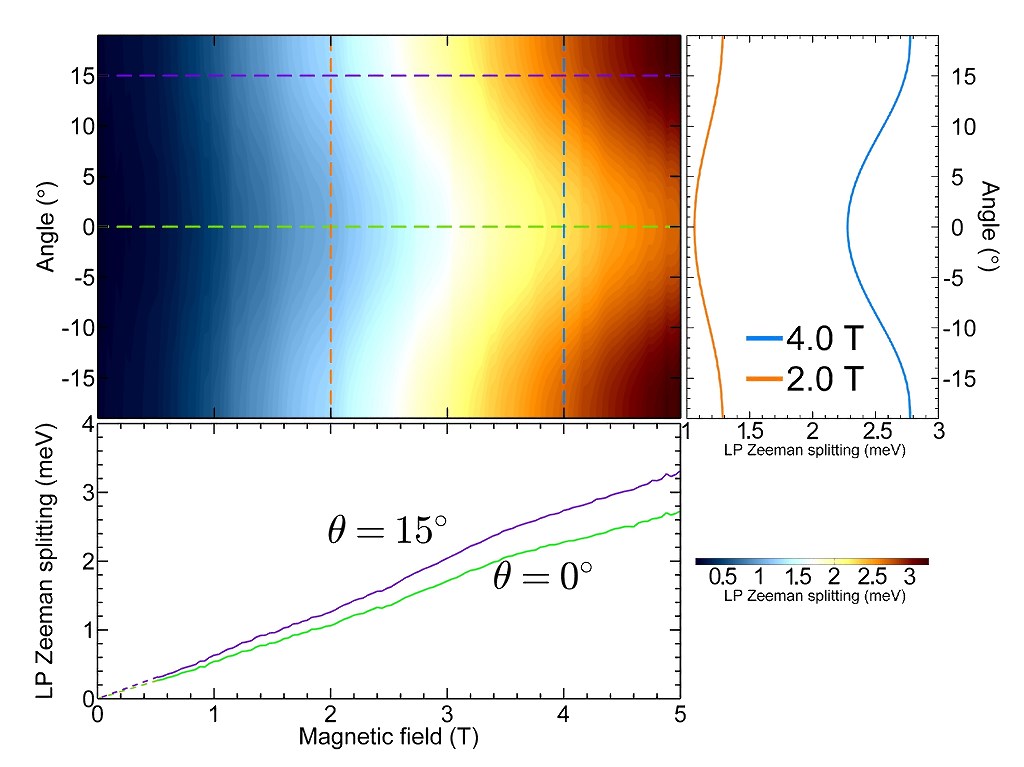}
 \caption{Map of the lower polariton Zeeman splitting in magnetic field for the results in positive detuning, $\delta_{0\textrm{T}}=+1.3$~meV. The right panel gives cross-sections of the map for different values of magnetic field. The bottom panel shows cross-sections for different angles of emission. The map color code illustrates the value of the Zeeman splitting of the lower polariton in meV.}
 \label{fig:fig_5}
 \end{figure}

The giant Zeeman effect of semimagnetic exciton-polaritons in magnetic field is illustrated in \hyperref[fig:fig_5]{FIG.~\ref{fig:fig_5}}. Based on the fitting of the solution of the Hamiltonian (1) to the experimental data for positive detuning illustrated in \hyperref[fig:fig_2b_skala]{FIG.~\ref{fig:fig_2b_skala}}, we plot the energy difference between the two magnetically split lower polariton branches for different emission angles. The right and bottom panels show cross-sections of the polariton Zeeman splitting for two different emission angles and for two different values of magnetic
field corresponding to the dashed lines marked on the map.

We observe that the lower polariton Zeeman splitting increases with emission angle, what indicates that the Zeeman splitting is higher for higher wavevectors. This effect results from the small effective mass of a microcavity photon compared to the one of a QW exciton. Therefore, independently of the detuning at zero wavevector, at higher wavevectors the upper polariton state is more photon-like, whereas the lower polariton state is more exciton-like. Consequently, the lower polariton state is significantly more affected by magnetic field \cite{Pietka_PRB2015, Walker_PRL2011} at high wavevectors and the lower polariton energy splitting due to $s$,$p$-$d$ exchange interaction is higher for high emission angles.

In the most general case, considering the coupling of cavity photonic mode, $E_{ph}$, to the heavy hole exciton in the QW, $E_{exc}$ and neglecting the coupling to the light hole exciton, the lower polariton energy in $\sigma^+$ and $\sigma^-$ polarizations can be expressed by the Hopfield coefficients in the following way:
\begin{align*}
	LP_{\sigma\pm}=\chi^2_{\sigma\pm}E_{exc}^{\sigma\pm}+C^2_{\sigma\pm}E_{ph}-\hbar\Omega\chi_{\sigma\pm}C_{\sigma\pm}
\end{align*}
where the $\chi^2$ and $C^2$ are the excitonic and photonic Hopfield coefficients, respectively \cite{hopfield}. These parameters give a direct measure of the excitonic and photonic content in the polariton state. The polariton Zeeman splitting, being the energy difference between $LP_{\sigma +}$ and $LP_{\sigma -}$, is therefore given by\\
	\begin{align*}
		\Delta E_{LP}&=\chi^2_{\sigma-}E_{exc}^{\sigma-}-\chi^2_{\sigma+}E_{exc}^{\sigma+}+(1-\chi^2_{\sigma-})E_{ph}-\\
 &\qquad-(1-\chi^2_{\sigma+})E_{ph}-\hbar\Omega(\chi_{\sigma-}C_{\sigma-}-\chi_{\sigma+}C_{\sigma+})
	\end{align*}	\\

or alternatively:\\

	\begin{align*}
	\Delta E_{LP}&=\chi^2_{\sigma-}(E_{exc}^{\sigma-}-E_{ph})-\chi^2_{\sigma+}(E_{exc}^{\sigma+}-E_{ph})+\\
	&\qquad+\hbar\Omega(\chi_{\sigma+}C_{\sigma+}-\chi_{\sigma-}C_{\sigma-})
	\end{align*}	\\

The polariton Zeeman splitting is therefore not directly proportional to the excitonic content in polariton state, as can be well approximated for manganese free GaAs-based structures \cite{Pietka_PRB2015}. The splitting is dependent on all three parameters (all influenced by the magnetic field): Hopfield coefficients, the photon-exciton detuning and the Rabi coupling. In the case of polaritons composed of semimagnetic excitons, the excitonic Hopfield coefficient is significantly influenced by magnetic field and the behaviour is opposite for $\sigma^+$ and $\sigma^-$ polarizations, what is illustrated in FIG.~6 a) for zero and high emission angles for 2~T and 4~T illustrated in FIG.~6~b).

The linewidth in CdMnTe-based structures is broader than in non-magnetic samples due the inhomogenous Mn distribution. At zero magnetic field it is of 2.35~meV, but it decreases for $\sigma^+$ to 2.2~meV at 5~T and increases for $\sigma^-$ polarization to 2.5~meV. (The mechanism for narrowing and broadening of the polariton emission linewidth in magnetic field is described in SI.) The value of polariton Zeeman splitting was approx.~3~meV at 5~T, at positive detuning at the temperatures of 10~K. However, the polariton Zeeman splitting in our structure can be as high as 10~meV at 5~T at sufficiently low temperature of 1.4~K, being much larger than polariton linewidth.
Nevertheless, even with the lower polariton linewidths comparable to the Zeeman splitting at 10~K, separation of the emission lines with opposite circular polarizations is observable in non-polarization-resolved experiment.

This is in contrast to the non-magnetic GaInAs structures. Even though, the polariton Zeeman splitting in GaInAs samples is of the order 0.1~meV at 5~T, being comparable or smaller than the polariton linewidth (approx. 0.2~meV) \cite{Pietka_PRB2015, Walker_PRL2011}, the significant diamagnetic shift in this materials always results in the decrease of the exciton content for the lower polariton branch at high magnetic fields, which ends up in the decrease of the observable polariton Zeeman splitting in magnetic field. Therefore even at magnetic field of 14~T and at positive detuning, the polariton Zeeman splitting is not resolved from polariton linewidth.

\begin{figure}
	\centering
	\includegraphics[width=\textwidth/2]{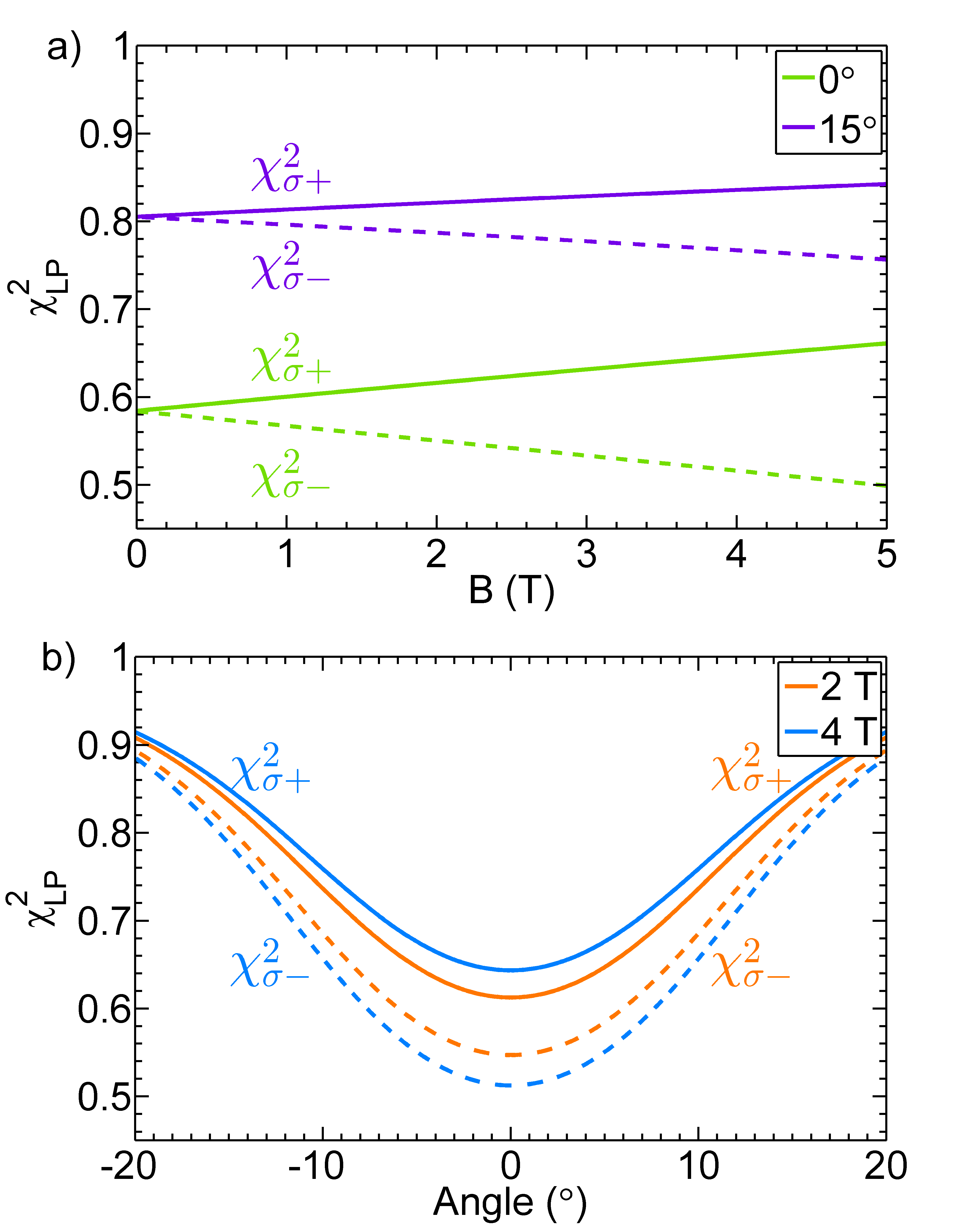}
	\caption{a) Evolution of the excitonic Hopfield coefficient, $\chi^2$, in magnetic field for $\sigma^+$ and $\sigma^-$ polarized polaritons and for two different emission angles calculated for the detuning of $\delta_{0\textrm{T}}=+1.3$~meV at 0~T. b) Angular dependence of $\chi^2$ for $\sigma^+$ and $\sigma^-$ polarized polaritons at magnetic field of 2~T and 4~T. }
	\label{fig:fig_6}
\end{figure}

Since for CdMnTe at low fields diamagnetic shift is negligible comparing to giant Zeeman effect, the energy shift in magnetic field for $\sigma^+$ excitons lead directly to increase of excitonic content ($\chi^2_{\sigma+}$) in polariton mode, as presented in \hyperref[fig:fig_6]{FIG.~\ref{fig:fig_6}~a)}.

The value of the observed polariton Zeeman splitting together with the absence of diamagnetic shift is remarkable and distinguishes our structure based on semimagnetic CdMnTe semiconductors from any other non-magnetic microcavity sample. Moreover, the value of the Zeeman splitting can be even further increased with higher Mn content. The size of this effect is a direct consequence of the exchange interaction between delocalized $s$- and $p$- type band electrons and localized $d$-type manganese ions. Even if Zeeman splitting follows a similar law for non-magnetic and magnetic samples its origin is different - for non-magnetic sample the splitting is caused by magnetic field which acts on both (i.e orbital and spin) parts of wavefunction of electron and hole, while in magnetic samples the splitting is due to exchange interaction, i.e. " exchange magnetic field " which acts only on spin part of wavefunction. This means, that magnetic sample is not simply a non-magnetic sample at very high magnetic field.

  \section{Summary}
We studied magnetic field effects on exciton-polaritons created in microcavities with semi-magnetic quantum wells. The manganese ions incorporated into quantum wells enhance magnetic field effects by $s$,$p$-$d$ exchange interaction. We demonstrated that the exciton-polaritons formed from semimagnetic excitons exhibit giant Zeeman splitting in magnetic field. We have shown that in these structures, magnetic field changes only the excitonic part of polaritons, and the splitting strongly depends on the photon-exciton detuning and polariton in-plane wavevector. Our results pave the way to the observation of many spin-related phenomena of exciton-polaritons in magnetic field. The giant Zeeman splitting can be crucial in the observation of the spin-Meissner effect
\cite{Walker_PRL2011}, and spin-dependent polariton-polariton interactions \cite{Paraiso-NM2010}. Our results and the fabricated structure are the first step to explore complex phenomena in non-equilibrium spinor condensates of exciton-polaritons based on semimagnetic structures.

\section{Acknowledgements}
This work was supported by the Polish National Science Centre under projects UMO-2014/13/N/ST3/03763, UMO-2015/16/T/ST3/00506, UMO-2015/18/E/ST3/00558, UMO-2015/18/E/ST3/00559, UMO-2013/09/B/ST3/02603 and UMO-2015/17/B/ST3/02273, and by the Polish Ministry of Science and Higher Education as research grants Iuventus Plus IP2014 040473 and IP2014 034573 in years 2015-2017. This study was carried out with the use of CePT, CeZaMat and NLTK infrastructures financed by the European Union - the European Regional Development Fund. Scientific work co-financed from the Ministry of Higher Education budget for education as a research project "Diamentowy Grant": 0010/DIA/2016/45 and 0109/DIA/2015/44.


\end{document}